\documentclass[journal]{IEEEtran}
\IEEEoverridecommandlockouts
\usepackage{psfrag}
\usepackage{cite}
\usepackage{amsmath,amssymb,amsfonts,mathbbol,acronym}
\usepackage[ruled,vlined]{algorithm2e}
\usepackage{algorithmic}
\usepackage{graphicx}
\usepackage{textcomp}
\usepackage[table]{xcolor}
\usepackage{bbm}
\usepackage{tabularx}
\usepackage{subfig}
\usepackage{cancel}
\usepackage{hyperref}
\usepackage{tikz}
\usepackage{pgfplots}
\usepackage{xparse}
\usepackage{soul}
\usepackage{comment}
\usepackage{colortbl,multirow,multicol,booktabs}
\usepackage{geometry}
\definecolor{LightCyan}{rgb}{0.88,1,1}
\pgfplotsset{compat=1.15}

\acrodef{5G}{fifth generation}
\acrodef{6G}{sixth generation}
\acrodef{AMIX}{Advantage-Function MIXing network}
\acrodef{AOA}{angle-of-arrival}
\acrodef{A3C}{Asynchronous Advantage Actor Critic}

\acrodef{CS}{conditioned stimulus}
\acrodef{CR}{conditioned response}
\acrodef{CAL}{Continuous Advantage Learning}
\acrodef{CRLB}{Cramer-Rao Lower Bound}
\acrodef{DAPG}{Demo-Augmented Policy Gradient}
\acrodef{DNN}{deep neural network}
\acrodef{DP}{dynamic programming}
\acrodef{DQL}{Deep $Q$-Learning}
\acrodef{DQN}{Deep $Q$-Network}
\acrodef{FIM}{Fisher Information Matrix}
\acrodef{GT}{goal-tracker}
\acrodef{GPS}{Global Positioning System}
\acrodef{LOS}{line-of-sight}
\acrodef{LSTM}{long short-term memory}
\acrodef{MADRL}{multi-agent deep reinforcement learning}
\acrodef{MARL}{multi-agent reinforcement learning}
\acrodef{MDP}{Markov Decision Process}
\acrodef{MPC}{Model predictive control}
\acrodef{MLE}{maximum likelihood estimator}
\acrodef{NS}{neutral stimulus}
\acrodef{NLOS}{non-line-of-sight}
\acrodef{PIT}{Pavlovian-Instrumental Transfer}
\acrodef{PAL}{Pavlovian Avoidance Learning}
\acrodef{PEB}{Position Error Bound}
\acrodef{RF}{radio-frequency}
\acrodef{RL}{reinforcement learning}
\acrodef{RPE}{reward prediction error}
\acrodef{RSS}{received signal strenght}

\acrodef{SPE}{state prediction errors}
\acrodef{ST}{sign-tracker}
\acrodef{SNR}{signal-to-noise ration}
\acrodef{TD}{temporal difference}
\acrodef{ToA}{time-of-arrival}
\acrodef{UAV}{unmanned aerial vehicle}
\acrodef{UR}{unconditioned response}
\acrodef{US}{unconditioned stimulus}

\acrodef{V2X}{vehicle-to-everything}

\newcommand{\p}{\mathbf{\p}}

\begin{document}
\bstctlcite{IEEEexample:BSTcontrol}

\title{A Cognitive Framework for Autonomous Agents: Toward Human-Inspired Design}

\author{
Francesco Guidi,~\IEEEmembership{Member,~IEEE}, 
Jingfeng Shan,~\IEEEmembership{Member,~IEEE},
Mehrdad Saeidi,\\
Enrico Testi,~\IEEEmembership{Member,~IEEE}, Elia Favarelli,~\IEEEmembership{Member,~IEEE}, Andrea Giorgetti,~\IEEEmembership{Senior Member,~IEEE},  Davide~Dardari,~\IEEEmembership{Fellow,~IEEE}, Alberto Zanella,~\IEEEmembership{Senior Member,~IEEE}, Giorgio Li Pira, \\
Francesca Starita, and Anna Guerra,~\IEEEmembership{Member,~IEEE}
\thanks{F. Guidi and A. Zanella are with the National Research Council of Italy (CNR), viale del Risorgimento 2, 40136 Bologna, Italy, e-mail:  \{francesco.guidi,alberto.zanella\}@cnr.it.\\
J. Shan, M. Saeidi, E. Testi, E. Favarelli, A. Giorgetti, D. Dardari and A. Guerra are with WiLab, University of Bologna, via dell'Universit\'a 50, 47521 Cesena, Italy, e-mail: \{enrico.testi, elia.favarelli,andrea.giorgetti, davide.dardari,anna.guerra3\}@unibo.it. \\
G. Li Pira and F. Starita are with the University of Bologna, Department of Psychology ``Renzo Canestrari", 47521 Cesena, Italy.
}
\thanks{This work was partially supported by the European Union under the Italian National Recovery and Resilience Plan (NRRP) of  NextGenerationEU (Mision 4  – Component 2  -Investment 1.1) Prin 2022 (No. 104, 2/2/2022, CUP J53C24002790006), under ERC Grant no. 101116257 (project CUE-GO: Contextual Radio Cues for Enhancing Decision Making in Networks of Autonomous Agents).}}
\maketitle



\begin{abstract}
This work introduces a human-inspired \ac{RL} architecture that integrates Pavlovian and instrumental processes to enhance decision-making in autonomous systems. While existing engineering solutions rely almost exclusively on instrumental learning, neuroscience shows that humans use Pavlovian associations to leverage predictive cues to bias behavior before outcomes occur. We translate this dual-system mechanism into a cue-guided \ac{RL} framework in which \ac{RF} stimuli act as conditioned (Pavlovian) cues that modulate action selection. The proposed architecture combines Pavlovian values with instrumental policy optimization, improving navigation efficiency and cooperative behavior in unknown, partially observable environments. Simulation results demonstrate that cue-driven agents adapt faster, achieving superior performance compared to traditional instrumental-solo agents. This work highlights the potential of human learning principles to reshape digital agents intelligence.
\end{abstract}

\begin{IEEEkeywords}
Cognitive UAVs, Reinforcement Learning, Pavlovian-instrumental Transfer, Multi-Agent Systems
\end{IEEEkeywords}

\acresetall
\section{Introduction}

Among the emerging technologies, autonomous systems have been identified as part of the breakthrough innovations expected in the next years \cite{EuropeanCommissionJRC2025}. Multi-agent systems have attracted increasing attention for sensing and exploration applications due to their robustness to single-point failures and their ability to gather information from diverse perspectives and time scales \cite{Jaiswal2022}. Despite these advantages, the development of efficient task completion remains a major challenge \cite{Dorigo2021}. In this regard, several \ac{RL} approaches have been proposed to ensure reliable real-time task completion \cite{Wen2020,Wu2021}.  
Traditional autonomous agent design has drawn inspiration from biological systems to design collective decision-making mechanisms, particularly animal behavior  such as insect colonies \cite{Na2022}. However, human cognition has been largely overlooked in engineering applications \cite{Hassabis2017}. 

In humans, instrumental and Pavlovian processes coexist and compete for behavior control, and their interaction has been formalized within modern \ac{RL} frameworks \cite{dayan2002reward}. Instrumental learning determines actions that maximize rewards or minimize punishments, whereas Pavlovian learning associates environmental stimuli/cues with desirable or aversive outcomes \cite{dayan2002reward}. Notably, Pavlovian responses differ from those governed by instrumental conditioning, because the responses are elicited reflexively by predictive cues and take place before the presentation of the outcome, i.e., of the \ac{US}. A classic example is Pavlov's experiment, where dogs began to salivate upon hearing a bell ring previously associated with food delivery.

\begin{table*}[t]
\centering
\caption{Examples of Classical (Pavlovian) Conditioning in Human Behavior}
\label{tab:classicalconditioningexamples}
\renewcommand{\arraystretch}{1.25}
\setlength{\tabcolsep}{3pt}
\begin{tabular}{p{2.6cm} p{3.3cm} p{3.3cm} p{3.3cm}}
\toprule
\textbf{Scenario} & \textbf{Neutral Stimulus (NS)} & \textbf{Unconditioned Stimulus (US)} & \textbf{Conditioned Stimulus (CS)} \\
\midrule
\rowcolor{white}
\textbf{Food Anticipation} & Sound of a bell before meals & Smell or taste of food (elicits salivation) & Bell sound alone elicits salivation \\
\rowcolor{cyan!8}
\textbf{Spatial Navigation Learning} & Particular room lighting cue & Finding food or goal object (elicits approach behavior) & Light cue alone directs movement toward expected reward \\
\rowcolor{white}
\textbf{Social Conditioning} & Gesture & Verbal praise or criticism
(elicits emotional response) & Gesture alone modulates emotional response \\
\rowcolor{cyan!8}
\textbf{Advertising} & Brand logo or jingle & Pleasant imagery or reward (elicits positive emotion) & Logo alone evokes positive affect \\
\bottomrule
\end{tabular}
\end{table*}

In engineering, Pavlovian and instrumental learning have often been merged or, in most cases, Pavlovian learning has been neglected, rather than being considered alongside instrumental learning, as in human cognition \cite{HayEtAl:J2}. Early bio-inspired architectures, such as hierarchical affective controllers \cite{lee2019decision} and cognitive radar systems \cite{haykin2006cognitive, haykin2012cognitive}, emulate perception–action cycles but do not fully incorporate cue-guided behavior. More recent frameworks, including \emph{MaxPain} \ac{RL} algorithms \cite{mahajan2024balancing, wang2021modular}, integrate Pavlovian value signals to bias action selection, whereas multi-agent and dual-critic actor–critic methods \cite{zhang2024cognitive,chen2024double} employ separate evaluators to stabilize learning. These approaches mimic the interaction between Pavlovian and instrumental learning systems observed in biological agents \cite{dayan2002reward}, as exemplified by the \ac{PIT} effect. A well-known effect in the neuroscientific literature that demonstrates how Pavlovian cues can bias instrumental behavior in humans and non-humans animals.

Inspired by these insights, this work investigates how cue-guided human \ac{RL} principles can enhance the decision-making of autonomous agents when navigating uncertain environments. 
In this sense, traditional model-based methods
struggle in complex or partially known environments, highlighting the need for learning-based methods \cite{Meyer2015,Guerra2020a,Guerra2020b}.  
Thus, machine learning techniques, particularly \ac{RL} and its deep variants, have become essential for adaptive navigation and environmental awareness \cite{Testi2020,Guerra2021,fontanesi2025deep}. Advanced strategies, including incremental curriculum learning, \ac{LSTM} networks, and transfer learning, further accelerate adaptation in dynamic environments \cite{Hodge2021}.

Building on prior work in bio-inspired control and multi-agent coordination \cite{lee2019decision, haykin2006cognitive, haykin2012cognitive, elfwing2017parallel, mahajan2024balancing, wang2021modular, campo2021collective, zhang2024cognitive, chen2024double}, we propose a new human-inspired architecture that integrates Pavlovian and instrumental model-free \ac{RL} with model-based planning processes. This architecture leverages \ac{RL} guided by radio-frequency cues to improve the decision-making of autonomous agents. While visual cues can enhance navigation, their reliability decreases under poor visibility, whereas radio-frequency cues may provide predictive signals to improve learning efficiency \cite{Kalia2008}.
More specifically, we aim to address the following research question: (i) {\em Can Pavlovian learning, through the exploitation of radio cues, effectively accelerate learning when combined with instrumental learning?} (ii) {\em Can a hybrid approach, merging model-based and -free learning processes, improve agents' decision-making?} 

To illustrate its potential, in Sec.\ref{sec:pavlovhumans} and Sec.~\ref{sec:learningaa} we describe Pavlovian learning in humans and how traditional learning in autonomous agents is designed. Then, we provide a novel human inspired framework in Sec.~\ref{sec:newframework}, with a case study demonstrating the effectiveness of the proposed framework in Sec.~\ref{sec:nresults}. Conclusions and future perspective are finally drawn in Sec.~\ref{sec:conclusions}.

\vspace{-0.2cm}

\section{Pavlovian Learning in Humans}
\label{sec:pavlovhumans}
Pavlovian and instrumental conditioning are foundational learning processes that shape human behavior through distinct but complementary associative mechanisms. At the neural level, Pavlovian control in both humans and animals primarily engages subcortical and limbic circuits, such as the amygdala, insular cortex, and nucleus accumbens shell, whereas instrumental control relies on cortico-striatal networks supporting goal-directed and habit-based learning, including the dorsal striatum and orbitofrontal cortex.
Understanding the differences between these forms of learning provides insight into how organisms adapt to their environments and offers a principled foundation for human-inspired designs of autonomous agents. 
\paragraph{Pavlovian Conditioning} 
Pavlovian conditioning involves learning predictive associations between environmental states or stimuli and biologically relevant outcomes, i.e., the so called \ac{US}. A \ac{NS} becomes a \ac{CS} when repeatedly paired with an \ac{US} that elicits an \ac{UR}. After learning, the \ac{CS} triggers a \ac{CR} even without the \ac{US}. Because the \ac{US} occurs independently of the organism’s actions, Pavlovian responses are considered stimulus-driven and anticipatory. Classic examples include preparatory reactions to cues predicting food, such as salivation, or threat, such as freezing behavior \cite{dayan2002reward}.
Pavlovian responses are generally involuntary and evoked directly by predictive cues.

\begin{table*}[t]
\centering
\caption{Common dictionary between autonomous agents and humans.}
\label{tab:parallel_learning}
\renewcommand{\arraystretch}{1.15}
\setlength{\tabcolsep}{4pt}

\newcommand{\partialrowcolor}{\cellcolor{cyan!5}}
\newcommand{\partialrowcolortwo}{\cellcolor{cyan!10}}

\begin{tabular}{p{3.0cm} p{6.2cm} p{6.2cm}}
\toprule
\partialrowcolortwo \textbf{Category} &
\partialrowcolortwo \textbf{Humans} &
\partialrowcolortwo \textbf{Autonomous (Digital) Agents} \\
\midrule

\textbf{Learning} &
Enduring change in behavior or response following experience. &
Inferring a policy from observations and reward signals. \\

\partialrowcolor \textbf{Reinforcement} &
\partialrowcolor Reinforcement arises when reward delivery or omission influences human responding; outcomes drive learning in Pavlovian and instrumental systems. &
\partialrowcolor Reinforcement occurs when a reward signal evaluates an action and updates value functions or policies through trial and error. \\

\textbf{Stimulus} &
Environmental event (cue) that alters behavior via Pavlovian (e.g., the \ac{CS}) or instrumental learning. &
Observations of the environment (e.g., in POMDPs), used as state information. \\

\partialrowcolor \textbf{Outcome} &
\partialrowcolor Biologically relevant event that elicits changes in behavior without the need of any learning. It is the reward/runishment (instrumental) or the \ac{US} (Pavlovian). &
\partialrowcolor Reward or punishment observed by the agent and used to update policies or value functions; internal state does not change unless explicitly modeled. \\

\textbf{Response} &
Behavioral responses: motor reflex (Pavlovian) or voluntary action (instrumental). Also includes physiological responses (autonomic arousal, hormones), subjective experience or thoughts &
Action executed by the agent (behavioral). \\

\partialrowcolor \textbf{Model-free RL} &
\partialrowcolor Habitual/model-free control: actions selected automatically due to prior reinforcement, not dependent on current outcome value, can persist even if the outcome becomes undesired. &
\partialrowcolor Decision-making based on cached value estimates (policy or action-value function); learning arises directly from interaction with the environment. \\

\textbf{Model-based RL} &
Goal-directed/model-based control: actions chosen based on predicted desirable outcomes. The outcome must be motivationally relevant at the moment of action &
Selecting actions via planning under a known model of the environment; internal model used to improve policy. \\

\bottomrule
\end{tabular}
\end{table*}

Conversely, instrumental conditioning involves selecting actions based on their consequences. Instrumental behaviors can be goal-directed, relying on action–outcome knowledge and sensitive to outcome revaluation, or habitual, where actions are triggered automatically by contextual cues. While habitual behaviors often operate independently of current goals (not sensitive to devaluation), they may still originate from previously goal-directed processes

Key components of Pavlovian conditioning include: $(i)$ \emph{Association formation}: repeated pairing of a \ac{NS} and \ac{US} produces a \ac{CR} to the \ac{NS}, which is now considered a \ac{CS}; $(ii)$ \emph{Biological preparedness}: \acp{US} naturally elicit \acp{UR} without the need for learning; and $(iii)$ \emph{Learned anticipation and value of the \ac{CS}}: \acp{CR} reflect the motivational and affective significance of the predicted outcome. Examples are provided in Table~\ref{tab:classicalconditioningexamples}.

\paragraph{Classes of Pavlovian Responses}
Pavlovian conditioning enables cues to acquire motivational value and elicit involuntary or reflexive behavioral or physiological responses, a process often described as value transfer from outcome to cue~\cite{starita2022pavlovian,delamater2007learning}. These responses adapt rapidly to outcome revaluation~\cite{pool2019behavioural}, highlighting that Pavlovian learning is not purely habitual or model-free~\cite{dayan2014model}.

Two major classes of \acp{CR} are typically distinguished~\cite{dayan2002reward}: \emph{preparatory responses}, involving autonomic or motivational adjustments (e.g., arousal changes), and \emph{consummatory responses}, involving outcome-specific motor patterns (e.g., approach or withdrawal). These responses may co-occur and rely on distinct neural systems. Pavlovian associations encode multiple outcome attributes \cite{dalbagno2025learning}, including sensory, motivational, and temporal features~\cite{pool2019behavioural}.

These principles motivate the integration of Pavlovian and instrumental mechanisms in autonomous agents to enable rapid, cue-guided, and context-sensitive behavior.
To facilitate the integration of human and digital worlds, Table~\ref{tab:parallel_learning} introduces a common dictionary defining key terms, such as learning, reinforcement, and model-free and model-based \ac{RL}, aimed also at supporting the  understanding of the following sections.

\section{Learning in Autonomous Agents}
\label{sec:learningaa}
In order to connect human-inspired learning with traditional learning approaches employed with autonomous agents, we first review the core concepts of \ac{TD} learning, emphasizing actor–critic models and their connection to Pavlovian and instrumental learning.

\subsection{Model-Based Learning}

Model-based learning refers to approaches that rely on an explicit model, either analytically derived or learned from data, to support decision-making or control. This paradigm spans from classical optimal control to modern model-based \ac{RL}, depending on how the model is acquired and exploited. The literature is so vast that here we briefly report few insights.

Classically, model-based control assumes known system dynamics. Foundational tools such as \ac{DP} solve sequential decision problems via the Bellman equation \cite{sutton1998reinforcement}, while the linear–quadratic regulator provides closed-form solutions with stability guarantees for linear systems \cite{anderson2007optimal}. \Ac{MPC} generalizes these ideas by solving constrained optimization problems over a receding horizon using a known model, effectively handling nonlinearities, constraints, and time-varying objectives \cite{rawlings2009model}.

Tools like Kalman and particle filters support sequential estimation \cite{Meyer2015}, while projected gradient descent has been applied for position optimization \cite{Guerra2020a,Guerra2020b}. 

When the model is unknown, model-based \ac{RL} learns an internal representation of environment dynamics.\footnote{A \ac{RL} method is here considered model-based if it learns (or is provided with) a model of the environment, namely the transition dynamics and the reward function. This model is then used to simulate experience for planning, rather than relying solely on real environment interactions as in model-free \ac{RL}.} This model enables planning via simulated rollouts, improving sample efficiency and safety, especially for systems where real-world trials are costly, such as \acp{UAV}. Early methods like Dyna-$Q$ \cite{sutton1990dyna} and Prioritized Sweeping \cite{moore1993prioritized} introduced principles of integrated learning and planning. In particular, Dyna-$Q$ integrates planning into the learning progress by using the learned environment model to generate $K$ simulated experiences for every real interaction with the environment. 
Recent advances couple probabilistic models with \ac{MPC} \cite{janner2019mbpo}, or learn latent dynamics from pixels 
\cite{hafner2019learning
}. MuZero and its variants combine learned latent models with tree search, removing the need for explicit reward modeling \cite{schrittwieser2020mastering}.

Beyond reward maximization, many real-world applications, such as target localization or exploration, prioritize reducing uncertainty. Information-seeking control extends model-based \ac{RL} by selecting actions that maximize expected information gain about hidden variables. For example, \cite{guerra2022networks} demonstrates \ac{UAV} swarms coordinating trajectories to optimize information acquisition under time constraints.
\vspace{-0.2cm}
\subsection{Model-Free Learning}
Model-free learning refers to approaches that learn to make decisions or select actions directly from experience, without relying on an explicit model of the environment's dynamics. This paradigm spans from classical \ac{TD} methods to modern deep \ac{RL}, depending on how value functions or policies are represented and updated from observed rewards and transitions.

\paragraph{Temporal-Difference Learning} 
\ac{TD} learning estimates the value function $V_\pi(s)$ of a fixed policy $\pi$ through experience as
\begin{equation}
\! \! V_{\pi}(s_t) \leftarrow V_{\pi}(s_t) + \alpha \left[r_{t+1} + \gamma V_{\pi}(s_{t+1}) - V_{\pi}(s_t)\right],
\end{equation}
where $\alpha$ is the learning rate, $\gamma$ is the discount factor, and the term in brackets is the \ac{TD} error. TD thus enables incremental, data-efficient learning akin to Pavlovian conditioning, predicting expected outcomes without altering action value.

In neuroscience, the Rescorla–Wagner model, a cornerstone of Pavlovian conditioning, parallels \ac{TD} learning by updating associative strengths based on prediction error. Learning occurs only when outcomes deviate from expectations, consistent with the concept of surprise-driven learning in \ac{RL} \cite{sutton1998reinforcement,starita2023threat}.

\paragraph{SARSA and $Q$-learning} 
Both algorithms extend TD learning to control by estimating the action-value function $Q(s,a)$. SARSA updates $Q(s_t,a_t)$ using the next action actually taken (on-policy),  
\begin{align}
&\! Q(s_t,a_t) \leftarrow Q(s_t,a_t) \\ \nonumber 
&\phantom{Q_{\text{S}}(s_t,a_t) \leftarrow }
+ \alpha [r_{t+1}\! +\! \gamma Q(s_{t+1},a_{t+1}) \!- \!Q(s_t,a_t)],
\end{align}
while $Q$-learning uses the greedy action (off-policy) to update $Q(s_t,a_t)$ as follows
\begin{align}
&Q(s_t,a_t) \leftarrow Q(s_t,a_t)  \\
&\phantom{Q_{\text{QL}}(s_t,a_t) \leftarrow}+ \alpha [r_{t+1} \!+ \!\gamma \max_{a'} Q(s_{t+1},a')\! - \! Q(s_t,a_t)].\nonumber 
\end{align}
SARSA propagates value estimates conservatively, reflecting experienced trajectories, while $Q$-learning tends to converge faster but may overestimate in stochastic environments. 

\paragraph{Deep $Q$-Learning} 
When state or action spaces are large, \acp{DNN} approximate $Q(s,a;\theta)$, with parameters $\theta$. Similar to $Q$-learning, the update rule for \ac{DQN} is based on the Bellman equation \cite{mnih2015human}. {A key innovation in \ac{DQL} is the use of a separate target network with parameters $\theta^{-}$, which stabilizes training. Specifically, \ac{DQL} frames the update as a supervised learning problem. A target value $y_t$ is computed using the target network}
\begin{equation}
y_t = r_{t+1} + \gamma \max_{a'} Q(s_{t+1}, a'; \theta^{-}).
\end{equation}
{The network is then trained by minimizing the mean-squared error loss between the target and its prediction}
\begin{align}
\mathcal{L}(\theta)=\mathbb{E}[(y_t - Q(s_t,a_t;\theta))^2]\,.
\end{align} 
Extensions, such as Double Dueling \ac{DQN}, allow to mitigate overestimation bias and improve learning efficiency by decoupling action evaluation and selection while explicitly estimating state value and advantage functions. 

\begin{figure*}
    \centering
    \includegraphics[width=0.8\linewidth]{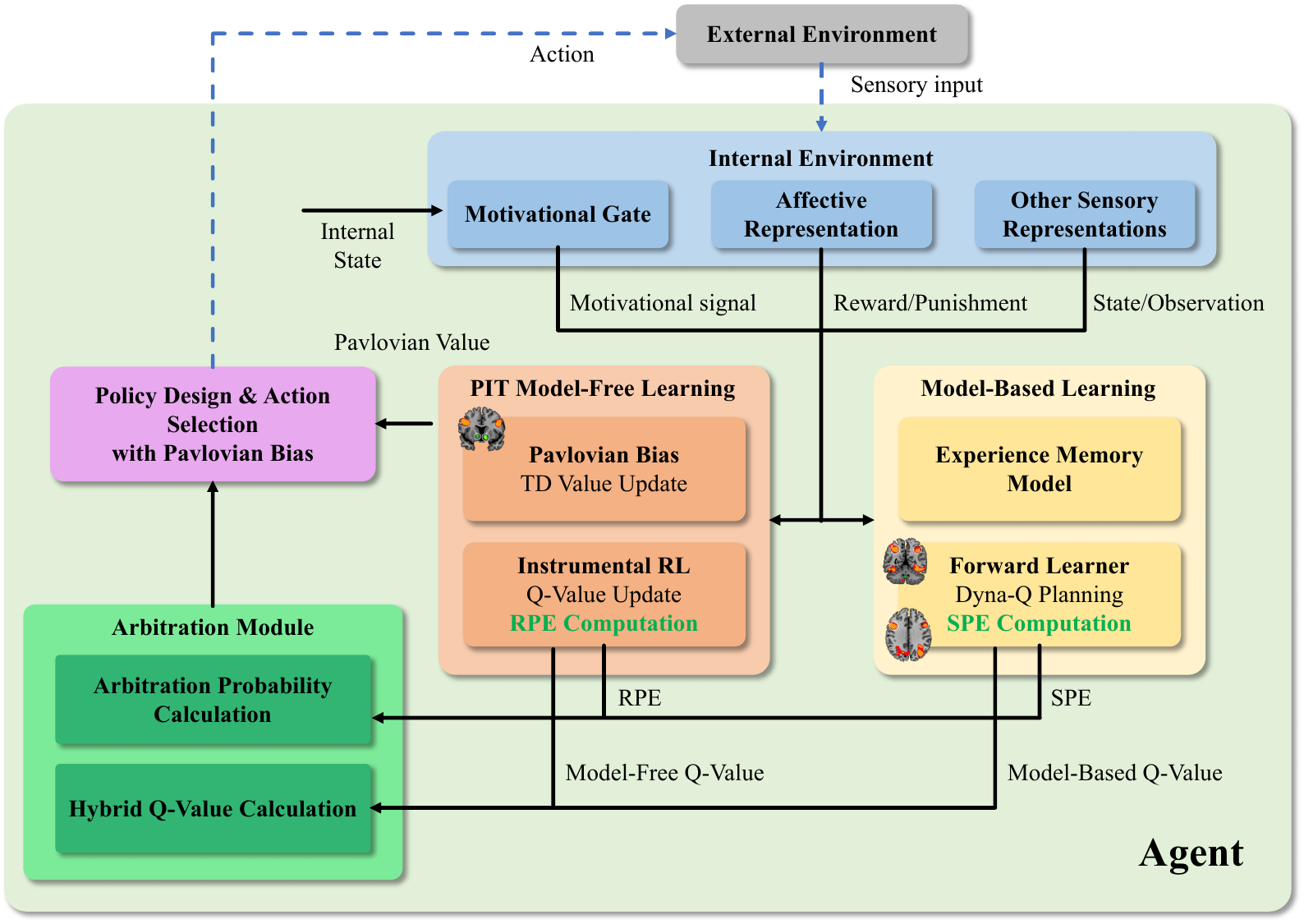}
    \caption{Proposed model-based and -free learning architecture inspired by \cite{lee2019decision}. The brain images illustrate neuroscientific correlates of the proposed mechanisms, adapted from \cite{o2015structure}: model-free learning is associated with neural activity correlating with reward prediction errors in the ventral striatum (shown in green), whereas model-based learning shows activity in posterior parietal cortex and dorsolateral prefrontal cortex correlating with state prediction errors (shown in orange).}
    \label{fig:diagram}
\end{figure*}


\subsection{Toward Human-Inspired Learning}
Actor-Critic architectures decompose learning into a \textit{critic}, which estimates the value function and computes the \ac{TD} error, and an \textit{actor}, which adjusts the policy to minimize that error. However, standard actor–critic models cannot fully explain Pavlovian behaviors, such as reflexive or motivation-driven responses that occur independently of instrumental control \cite{dayan2002reward}. 

In humans and animals, Pavlovian and instrumental learning do not act independently but interact to control behavior. To achieve a mathematical representation of their integration, the authors in \cite{dayan2002reward} extended actor–critic models with the \textit{advantage function}
\begin{equation}
A(s,a) = Q(s,a) - V(s),
\end{equation}
which measures how much better an action is than average. As learning progresses, advantages for suboptimal actions become negative and those for optimal actions vanish, modeling the behavioral shift from flexible, goal-directed learning to habitual or Pavlovian control.
Such an advantage function can be replicated in the design of learning approaches for digital agents in order to account for a Pavlovian component that allows to mimic biological behavior.
In addition, humans exhibit both deliberative (model-based) and habitual (model-free) learning systems \cite{
lee2014neural}. An arbitration mechanism regulates their influence according to reliability, quantified via prediction errors: \ac{RPE} for model-free learning and \ac{SPE} for model-based learning.


\section{A Novel Human-Inspired Architecture for Autonomous Agents}
\label{sec:newframework}
Pavlovian conditioning involves learning the association between a cue and its outcome. 
A radio-feature of the environment, initially interpreted as a \ac{NS},
can be associated to a cue that can elicit a certain behavior in order to accelerate the learning procedure.
To capture such a behavior, we propose an ad-hoc architecture which is reported in Fig.~\ref{fig:diagram}, that illustrates the interplay between an autonomous (digital) agent (e.g., an \ac{UAV}) and its environment. Through repeated interactions, the agent learns to act adaptively within a dynamic context. The environment includes both physical surroundings and social entities, such as other agents or humans.

In the following, we describe the proposed framework, constituted by an external perception–action loop and the internal learning mechanisms, drawing a parallel between digital agents and human cognitive systems.

\textbf{External Perception–Action Loop.}  
The perception–action loop represents the bidirectional interaction between the agent and its environment, where sensory inputs guide actions and actions modify subsequent sensory experiences. It is depicted with dashed blue lines in Fig.~\ref{fig:diagram}. Humans and digital agents share analogous mechanisms for processing perception and action.

\begin{table*}[h!]
\centering
\caption{Comparison between human and digital agents across functional modules.}
\label{tab:human_digital_comparison}
\renewcommand{\arraystretch}{1.25}
\begin{tabular}{p{0.22\linewidth} p{0.36\linewidth} p{0.36\linewidth}}
\toprule
\textbf{Module} & \textbf{Humans} & \textbf{Autonomous Agent} \\
\midrule

\textbf{Perception–Action Loop} &
Sensory inputs processed in cortical areas (e.g., visual, auditory, parietal) enable active perception, selective attention, and context-dependent action selection. &
Onboard sensors (e.g., cameras, antennas, LIDAR) collect data, which are processed to extract features guiding adaptive actions in a feedback loop. \\

\rowcolor{cyan!8}
\textbf{Internal Environment} &
Motivational drives emerge from homeostatic needs and emotions, regulated by neural systems such as the hypothalamus and VTA. Affective evaluation assigns subjective value to experiences. &
Internal states (e.g., energy, mission priority) influence a motivational gate that modulates learning sensitivity and balances external and internal reward signals. \\

\textbf{Pavlovian–instrumental (Model-Free)} &
The amygdala and ventral striatum learn cue–outcome associations. Dopaminergic RPE signals guide reflexive, habitual, and affective responses. &
Model-free RL associates states with affective values through reward prediction errors, shaping rapid, heuristic responses and habitual behavior. \\

\rowcolor{cyan!8}
\textbf{Instrumental Model-Based Learning} &
The hippocampus and prefrontal cortex simulate future scenarios, enabling flexible, goal-directed planning and cognitive control. &
Model-based RL learns and updates transition models to plan and evaluate long-term outcomes, supporting adaptive decision-making. \\

\textbf{Arbitration Mechanism} &
Cortical regions such as ACC and OFC monitor uncertainty and outcomes, balancing habitual and goal-directed control. &
Meta-controllers or probabilistic estimators dynamically weight model-free and model-based policies to trade off efficiency and flexibility. \\

\bottomrule
\end{tabular}
\end{table*}

\textbf{Internal Environment.}  
While perception and action are externally driven, their prioritization is influenced by the internal state, such as energy/battery level, time constraints, or motivational drives that regulate behavior. These internal processes emulate those found in biological organisms.

\textbf{Learning Systems.}  
The agent learns through two complementary systems: (1) a \textit{Pavlovian–Instrumental (P–I) Model-Free} subsystem, responsible for habitual and reflexive behavior, and (2) a \textit{Model-Based Reinforcement Learning} subsystem, enabling flexible and goal-directed control. 

\textbf{Pavlovian–instrumental (P–I) Model-Free.}  
This system integrates Pavlovian conditioning and model-free \ac{RL} through a shared value estimation process. In humans, the amygdala and ventral striatum evaluate cues predicting rewards or threats, while dopaminergic neurons in the VTA and substantia nigra encode reward prediction errors. In digital agents, this mechanism can be implemented as value-based learning, where states are associated with affective outcomes that bias action selection. This component provides rapid, emotion-like heuristics that shape reflexive responses and habits.

\textbf{Instrumental Model-Free Learning.}  
Humans acquire habits through repeated reinforcement, mediated by cortico-striatal circuits involving the prefrontal cortex and dorsomedial striatum. Similarly, digital agents use model-free algorithms such as $Q$-learning or SARSA to learn from experience without explicitly modeling state transitions. These systems are fast and efficient but may lack flexibility when conditions change.

\textbf{Instrumental Model-Based Learning.}  
Model-based learning enables flexible, goal-oriented behavior by using internal models of environmental dynamics. In humans, this relies on the hippocampus and prefrontal cortex to simulate future scenarios and evaluate potential outcomes. In digital agents, model-based \ac{RL} builds transition models from data, allowing planning and long-term optimization of actions.

\textbf{Arbitration Module.}  
The arbitration module regulates the contribution of model-free and model-based learning to action selection. Rather than committing to a single control strategy, the agent continuously evaluates the reliability of each subsystem based on their prediction errors. When the internal model provides accurate predictions, model-based planning is favored. Conversely, in uncertain modeled situations, control is shifted toward the more robust model-free system. More specifically, model-free and model-based $Q$-value estimates are combined into a hybrid action-value representation, whose balance is governed by an \emph{arbitration probability}. This probability reflects the relative confidence assigned to each learning subsystem and determines how strongly their respective value estimates influence the final decision policy \cite{geerts2020general}.

In humans, behavioral control is thought to arise from a similar arbitration process involving cortical regions such as the anterior cingulate cortex (ACC) and the orbitofrontal cortex (OFC), which monitor uncertainty, evaluate expected outcomes, and regulate the engagement of habitual versus goal-directed control~\cite{lee2014neural}. 
Similarly, from a computational perspective, arbitration relies on distinct prediction-error signals generated by the two learning systems. In the model-free case, \acp{RPE} encode discrepancies between expected and experienced outcomes~\cite{schultz1997neural, dabney2020distributional}. 
In contrast, model-based learning depends on \acp{SPE}, which quantify mismatches between predicted and observed state transitions and reflect the accuracy of the agent’s internal model~\cite{glascher2010states}.

By jointly monitoring \acp{RPE} and \acp{SPE}, the arbitration module estimates the evolving reliability of model-free and model-based subsystems and updates the arbitration probability and state-action values accordingly.

\textbf{Policy Design \& Action Selection.} 
Action selection is implemented through a softmax policy that jointly accounts for instrumental action values and Pavlovian biases. 
The Pavlovian term depends on both the current state and the selected action, enabling approach or avoidance tendencies toward specific regions of the environment. This action dependence is essential: a purely state-dependent Pavlovian value would affect all actions equally and would therefore have no influence on the resulting action probabilities after softmax normalization. 

In the hybrid setting, the instrumental action-value used by the policy is replaced by a hybrid value estimate that combines model-free and model-based information. The relative contribution of these two components is regulated by the arbitration mechanism.

Based on these considerations related to the proposed human-inspired architecture, 
Table~\ref{tab:human_digital_comparison} reports the comparison of the previously described functional modules of each system, illustrating how perception–action loops, internal states, Pavlovian and instrumental subsystems, and arbitration mechanisms operate in both biological and artificial learners. These mappings clarify how a human-inspired architecture enables digital agents to combine Pavlovian cue-driven responses, the stability of model-free learning, and the flexibility of model-based planning, improving navigation and decision-making.

\section{Simulation Results}
\label{sec:nresults}

To validate the proposed framework, we considered a team of autonomous agents that have to navigate an environment in order to localize a target with an error below a required \ac{PEB} threshold.
We assume that the environment is characterized by obstacles (shown in black), GPS-denied areas (in grey), and gates (in yellow) that allow agents to be in \ac{LOS} with the target. Gates and GPS-denied areas are treated as positive and negative cues, respectively.

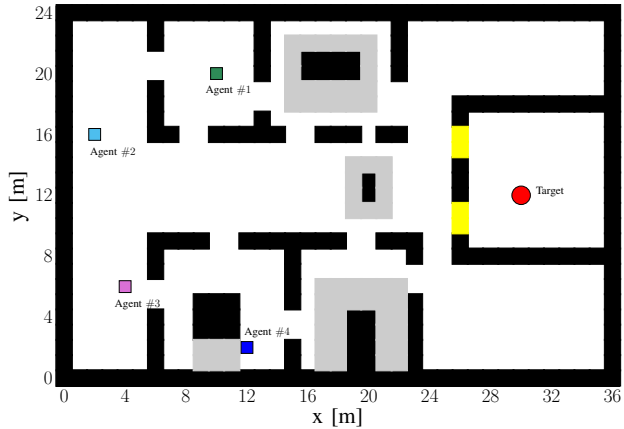
\begin{figure}[t!]
%
%
\definecolor{mycolor1}{rgb}{0.00000,0.44700,0.74100}%
\definecolor{mycolor2}{rgb}{0.85000,0.32500,0.09800}%
\definecolor{mycolor3}{rgb}{0.92900,0.69400,0.12500}%
\definecolor{mycolor4}{rgb}{1.00000,1.00000,0.00000}%
\definecolor{mycolor5}{rgb}{0,1,0}%
\definecolor{mycolor6}{rgb}{0.180, 0.545, 0.341}%
\definecolor{mycolor7}{rgb}{0.30196,0.74510,0.93333}%
\definecolor{mycolor8}{rgb}{0.49400,0.18400,0.55600}%
\definecolor{mycolor9}{rgb}{0.855, 0.439, 0.839}%
\definecolor{mycolor10}{rgb}{0.992, 0.737, 0.706}%
\definecolor{mycolor11}{rgb}{0, 0, 1}%

\begin{tikzpicture}[scale=0.35]
\tikzset{every node/.style={font=\large}}

\begin{axis}[%
width=8.427in,
height=5.694in,
at={(0.612in,0.864in)},
scale only axis,
xmin=-0.539055724707299,
xmax=36.4609442752927,
xtick={0,4,8,12,16,20,24,28,32,36},
tick label style={font=\huge},
xlabel style={font=\color{white!15!black}},
xlabel={x [m]}, 
xlabel style={font=\Huge},
ymin=-0.563745259817576,
ymax=24.4362547401824,
ytick={ 0,  4,  8, 12, 16, 20, 24, 28},
ylabel style={font=\color{white!15!black}},
ylabel={y [m]}, 
ylabel style={font=\Huge},
axis background/.style={fill=white}
]
\addplot [color=black, dashed, line width=0.1pt, forget plot]
  table[row sep=crcr]{%
-0.5	0\\
36.5	0\\
};
\addplot [color=black, dashed, line width=0.1pt, forget plot]
  table[row sep=crcr]{%
-0.5	24\\
36.5	24\\
};
\addplot [color=black, dashed, line width=0.1pt, forget plot]
  table[row sep=crcr]{%
0	-0.5\\
0	24.5\\
};
\addplot [color=black, dashed, line width=0.1pt, forget plot]
  table[row sep=crcr]{%
36	-0.5\\
36	24.5\\
};
\addplot [color=mycolor1, only marks, mark size=8.8pt, mark=square*, mark options={solid, fill=black, draw=black}, forget plot]
  table[row sep=crcr]{%
0	0\\
1	0\\
2	0\\
3	0\\
4	0\\
5	0\\
6	0\\
7	0\\
8	0\\
9	0\\
10	0\\
11	0\\
12	0\\
13	0\\
14	0\\
15	0\\
16	0\\
17	0\\
18	0\\
19	0\\
20	0\\
21	0\\
22	0\\
23	0\\
24	0\\
25	0\\
26	0\\
27	0\\
28	0\\
29	0\\
30	0\\
31	0\\
32	0\\
33	0\\
34	0\\
35	0\\
36	0\\
0	24\\
1	24\\
2	24\\
3	24\\
4	24\\
5	24\\
6	24\\
7	24\\
8	24\\
9	24\\
10	24\\
11	24\\
12	24\\
13	24\\
14	24\\
15	24\\
16	24\\
17	24\\
18	24\\
19	24\\
20	24\\
21	24\\
22	24\\
23	24\\
24	24\\
25	24\\
26	24\\
27	24\\
28	24\\
29	24\\
30	24\\
31	24\\
32	24\\
33	24\\
34	24\\
35	24\\
36	24\\
0	1\\
0	2\\
0	3\\
0	4\\
0	5\\
0	6\\
0	7\\
0	8\\
0	9\\
0	10\\
0	11\\
0	12\\
0	13\\
0	14\\
0	15\\
0	16\\
0	17\\
0	18\\
0	19\\
0	20\\
0	21\\
0	22\\
0	23\\
36	1\\
36	2\\
36	3\\
36	4\\
36	5\\
36	6\\
36	7\\
36	8\\
36	9\\
36	10\\
36	11\\
36	12\\
36	13\\
36	14\\
36	15\\
36	16\\
36	17\\
36	18\\
36	19\\
36	20\\
36	21\\
36	22\\
36	23\\
};
\addplot [color=mycolor2, only marks, mark size=8.8pt, mark=square*, mark options={solid, fill=black, draw=black}, forget plot]
  table[row sep=crcr]{%
6	1\\
15	1\\
19	1\\
20	1\\
23	1\\
6	2\\
15	2\\
19	2\\
20	2\\
23	2\\
6	3\\
9	3\\
10	3\\
11	3\\
19	3\\
20	3\\
23	3\\
6	4\\
9	4\\
10	4\\
11	4\\
19	4\\
20	4\\
23	4\\
9	5\\
10	5\\
11	5\\
15	5\\
23	5\\
15	6\\
6	7\\
15	7\\
6	8\\
15	8\\
23	8\\
26	8\\
27	8\\
28	8\\
29	8\\
30	8\\
31	8\\
32	8\\
33	8\\
34	8\\
35	8\\
6	9\\
7	9\\
8	9\\
9	9\\
12	9\\
13	9\\
14	9\\
15	9\\
16	9\\
17	9\\
18	9\\
21	9\\
22	9\\
23	9\\
26	9\\
20	12\\
26	12\\
20	13\\
26	13\\
26	14\\
6	16\\
7	16\\
10	16\\
11	16\\
12	16\\
13	16\\
14	16\\
17	16\\
18	16\\
19	16\\
21	16\\
22	16\\
6	17\\
13	17\\
26	17\\
6	18\\
26	18\\
27	18\\
28	18\\
29	18\\
30	18\\
31	18\\
32	18\\
33	18\\
34	18\\
35	18\\
6	19\\
13	20\\
16	20\\
17	20\\
18	20\\
19	20\\
22	20\\
13	21\\
16	21\\
17	21\\
18	21\\
19	21\\
22	21\\
6	22\\
13	22\\
22	22\\
6	23\\
13	23\\
22	23\\
};
\addplot [color=mycolor3, only marks, mark size=8.8pt, mark=square*, mark options={solid, fill=mycolor4, draw=mycolor4}, forget plot]
  table[row sep=crcr]{%
26	10\\
26	11\\
26	15\\
26	16\\
};
\addplot [color=white!80!black, only marks, mark size=8.8pt, mark=square*, mark options={solid, fill=white!80!black}, forget plot]
  table[row sep=crcr]{%
9	1\\
10	1\\
11	1\\
9	2\\
10	2\\
11	2\\
15	20\\
15	21\\
15	22\\
20	20\\
20	21\\
16	22\\
17	22\\
18	22\\
19	22\\
20	22\\
17	1\\
17	2\\
17	3\\
17	4\\
17	5\\
17	6\\
18	1\\
18	2\\
18	3\\
18	4\\
18	5\\
18	6\\
19	5\\
19	6\\
20	5\\
20	6\\
21	1\\
21	2\\
21	3\\
21	4\\
21	5\\
21	6\\
22	1\\
22	2\\
22	3\\
22	4\\
22	5\\
22	6\\
15	18\\
16	18\\
17	18\\
18	18\\
19	18\\
20	18\\
15	19\\
16	19\\
17	19\\
18	19\\
19	19\\
20	19\\
19	11\\
19	12\\
19	13\\
19	14\\
20	11\\
20	14\\
21	11\\
21	12\\
21	13\\
21	14\\
};
\addplot [color=mycolor6, dashed, mark size=10.0pt, mark=*, mark options={solid, fill=red, draw=black}, forget plot]
  table[row sep=crcr]{%
30	12\\
};
\node[right, align=left, inner sep=0]
at (axis cs:31.683,12.222) {$\!\!\!\!\!\!\!\!$ Target};
\addplot [color=mycolor6, only marks, mark size=6.4pt, mark=square*, mark options={solid, fill=mycolor6, draw=black}, forget plot]
  table[row sep=crcr]{%
10	20\\
};
\node[right, align=left, inner sep=0]
at (axis cs:9.286,18.927) {Agent $\#1$};
\addplot [color=mycolor7, only marks, mark size=6.4pt, mark=square*, mark options={solid, fill=mycolor7, draw=black}, forget plot]
  table[row sep=crcr]{%
2	16\\
};
\node[right, align=left, inner sep=0]
at (axis cs:1.686,14.796) {Agent $\#2$};
\addplot [color=mycolor9, only marks, mark size=6.4pt, mark=square*, mark options={solid, fill=mycolor9, draw=black}, forget plot]
  table[row sep=crcr]{%
4	6\\
};
\node[right, align=left, inner sep=0]
at (axis cs:3.313,4.853) {Agent $\#3$};
\addplot [color=mycolor11, only marks, mark size=6.4pt, mark=square*, mark options={solid, fill=mycolor11, draw=black}, forget plot]
  table[row sep=crcr]{%
12	2\\
};
\node[right, align=left, inner sep=0]
at (axis cs:11.825,3.022) {Agent $\#4$};
\end{axis}

\end{tikzpicture}%
    \caption{Simulated scenario. 
}
    \label{fig:simulatedscenario}
\end{figure}

\subsection{Simulated Scenario}

The simulated environment consists of a grid with $24 \times 36$ cells, where each cell has an area of $1\,\text{m}^2$ (see Fig.~\ref{fig:simulatedscenario}). Four autonomous agents operate within this space, starting at positions with coordinates $(20, 4)$, $(16, 2)$, $(6, 4)$, and $(2, 12) \, \text{m}$.
The target to be localized is placed at $(12, 30)$ and broadcasts periodic beacon signals used by the agents for localization.
\begin{figure}[t!]
\centering
\input{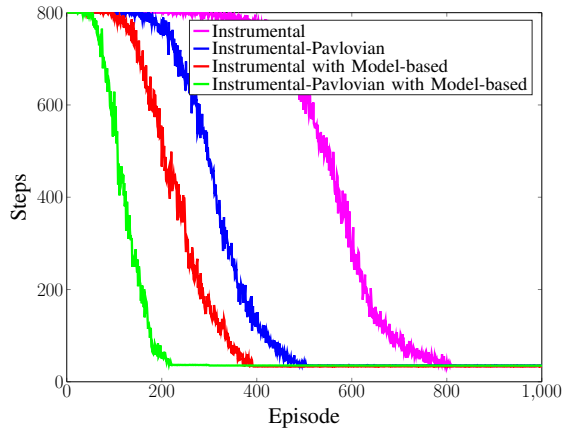}
    \caption{Learning rate expressed as the number of steps required to accomplish the mission per episode. 
}
    \label{fig:learningrate}
\end{figure}
Agents navigate the grid using a discrete set of five possible actions: moving up, down, left, or right, or hovering. Their movement is restricted by the environment’s boundaries, structural obstacles, and the presence of other agents. When conflicts arise, e.g., when two agents attempt to move into the same cell or cross paths simultaneously, these collisions are avoided by assigning penalties.

\begin{figure*}
    \centering
    \includegraphics[width=1\textwidth]{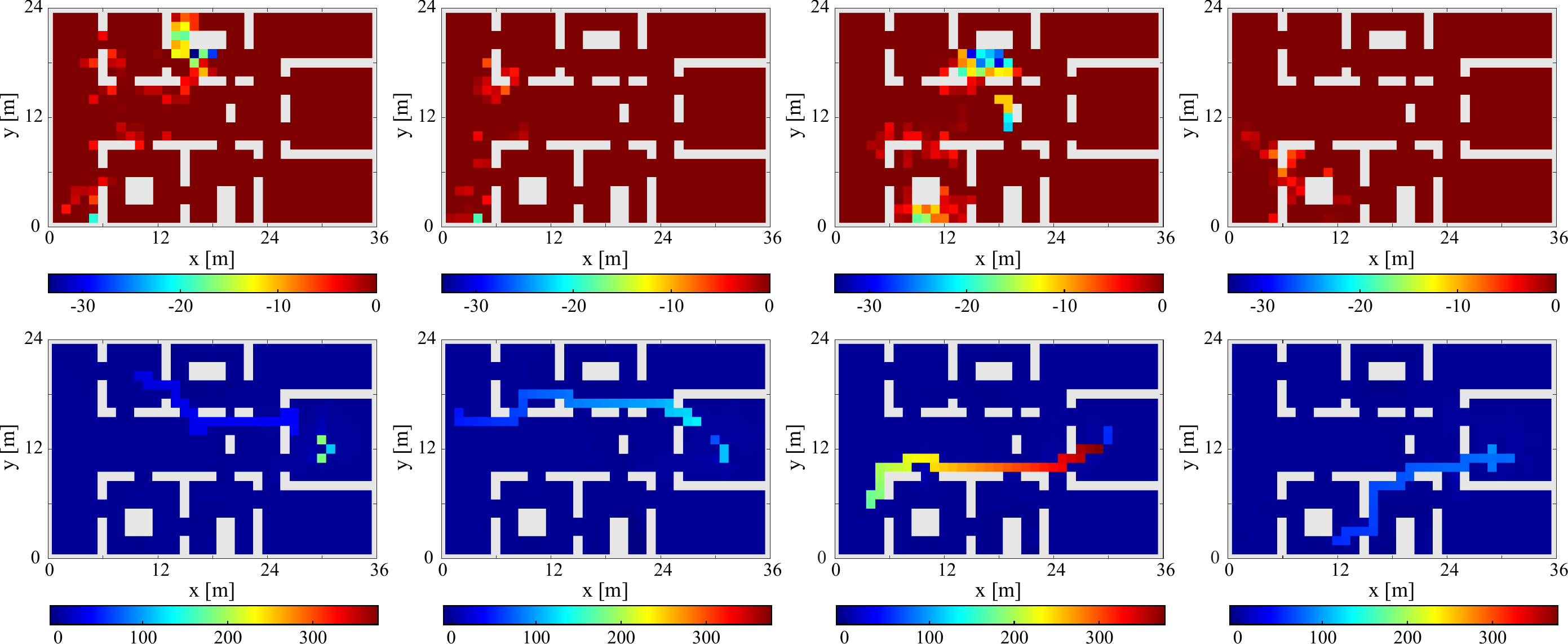}
    \caption{Pavlovian values at the end of episode 1 (top) and 400 (bottom), for agents \#1 (left) to \#4 (right), respectively.}
     \label{fig:pavlovianep1}
\end{figure*}

To sense the target, agents rely mainly on the \ac{RSS} of the beacon signal, with attenuation factor of $d^{\beta}$, where $d$ is the agent-target distance and $\beta$ is the path-loss exponent ($\beta=2$ and $\beta=3.5$ under \ac{LOS} and \ac{NLOS} conditions, respectively). 
The transmitted power is set to $-10\,$dBm, the TX/RX antenna gain to $2\,$dBi, the signal bandwidth to $1\,$MHz around a central frequency of $2.4\,$GHz and the receiver noise figure to $10\,$dB. Then, in areas where {GPS} is unavailable, the agents' position estimate is corrupted by a further random Gaussian error with variance of $100\,\text{m}^2$ for each spatial coordinate.

\subsection{Learning Parameters and Reward Structure}

Each agent is trained over $2400$ episodes, with each episode lasting up to $800$ time steps. Both the {off-policy Q-learning} and the Pavlovian critic rely on the same schedule for adjusting their learning rates and discount factors. In particular, their $\alpha$ start from $0.55$ and gradually decrease to $0.08$ with a decay factor of $0.999$, where $\gamma=0.99$. Action selection is governed by a softmax mechanism whose temperature likewise decays over time, starting at $1$, decreasing with a factor of $0.999$, and eventually reaching $0.08$.

The reward structure combines instrumental and Pavlovian components, and it is modelled according to the \ac{RSS}. Instrumental penalties include a cost of $2$ whenever an agent collides with an obstacle or another agent. Pavlovian influences are tied to environmental features: entering a designated “gate’’ region grants a positive reward of $5$, moving into GPS-denied areas triggers a negative reward of $5$, and operating under NLOS conditions imposes an additional penalty of $2$. These combined rewards shape both the deliberate and reflexive aspects of agent behavior during learning.

Regarding the model-based component, we considered a Dyna-$Q$ architecture with a planning step set to $4$, to balance the simulation efficiency and effectiveness. 

Regarding the Pavlovian reward values used in our proposed framework, they are derived from learned associations between environmental cues and task outcomes.
During an initial exploration stage, agents record the cumulative instrumental reward following exposure to each candidate cue. If a cue reliably predicts improved performance, such as higher localization accuracy and reduced path loss, it is assigned a positive Pavlovian value proportional to the average benefit. Conversely, the same holds for negative cues.

Cue association can be implemented using a TD learning rule, ensuring that Pavlovian rewards capture the motivational significance of each cue and enable cue-driven behavioral modulation without interfering with instrumental policy optimization. For example, \cite{SaeEtAl:C26} formalize cue–outcome association using a two-phase structure: a pre-cue phase, in which the agent applies standard off-policy $Q$-learning with environment rewards only, and a post-cue phase, triggered upon first encountering a validated cue. In the post-cue phase, potential-based reward shaping is applied to all remaining transitions, and learning switches to SARSA with replacing eligibility traces limited to the post-cue trajectory.

\subsection{Performance}

To quantitatively assess the impact of Pavlovian conditioning on learning performance, we ran parallel simulations under four conditions: (i) instrumental-only learning; (ii) Pavlovian–instrumental learning; (iii) instrumental-only learning with model-based planning; and (iv) instrumental-Pavlovian with model-based learning.
Figure~\ref{fig:learningrate} shows results over 60 Monte Carlo runs. The Pavlovian–instrumental agent reaches a competent policy far earlier than the instrumental-only baseline, confirming that cue-driven learning accelerates exploration toward more rewarding regions. By contrast, the instrumental-only learner progresses more slowly and exhibits higher variance across trials. Thus, Pavlovian cues supply an immediate potential that guides exploration toward perceptually informative areas and away from GPS-denied areas, reducing the search burden and enabling faster identification of valuable regions such as gates and \ac{LOS} corridors. The instrumental component then refines actions and long-horizon rewards for effective multi-agent localization.
Adding model-based planning further improves performance. Instrumental plus model-based learning yields better long-term efficiency through forward simulation, with Dyna-$Q$ outperforming pure instrumental learning but still suffering from a cold start as it builds its model. The full integration of Pavlovian, instrumental, and model-based learning performs best, combining early cue-driven guidance with later, accurate planning. This synergy yields the fastest convergence and more consistent identification of optimal paths.

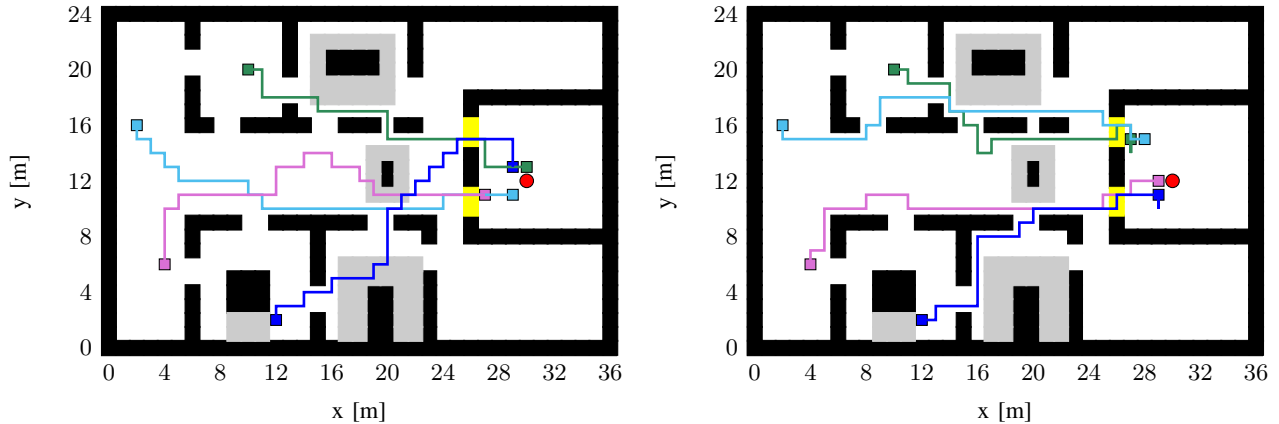
\begin{figure*}[t!]
\centering
\begin{minipage}[c]{0.48\textwidth}
\centering
%
%
\definecolor{mycolor1}{rgb}{0.00000,0.44700,0.74100}%
\definecolor{mycolor2}{rgb}{0.85000,0.32500,0.09800}%
\definecolor{mycolor3}{rgb}{0.92900,0.69400,0.12500}%
\definecolor{mycolor4}{rgb}{1.00000,1.00000,0.00000}%
\definecolor{mycolor5}{rgb}{0,1,0}%
\definecolor{mycolor6}{rgb}{0.180, 0.545, 0.341}%
\definecolor{mycolor7}{rgb}{0.30196,0.74510,0.93333}%
\definecolor{mycolor8}{rgb}{0.49400,0.18400,0.55600}%
\definecolor{mycolor9}{rgb}{0.855, 0.439, 0.839}%
\definecolor{mycolor10}{rgb}{0.992, 0.737, 0.706}%
\definecolor{mycolor11}{rgb}{0, 0, 1}%
\begin{tikzpicture}[scale=0.32]
\tikzset{every node/.style={font=\small}}
\pgfplotsset{set layers}

\begin{axis}[%
width=8.427in,
height=5.694in,
at={(0.612in,0.864in)},
scale only axis,
xmin=-0.539055724707299,
xmax=36.4609442752927,
xtick={0,4,8,12,16,20,24,28,32,36},
tick label style={font=\small},
xlabel style={font=\color{white!15!black}, yshift=-10pt},
xlabel={x [m]}, 
xlabel style={font=\small},
ymin=-0.563745259817576,
ymax=24.4362547401824,
ytick={ 0,  4,  8, 12, 16, 20, 24, 28},
ylabel style={font=\color{white!15!black}, yshift=18pt},
ylabel={y [m]}, 
ylabel style={font=\small},
axis background/.style={fill=white}
]
\addplot [color=mycolor1, only marks, mark size=8.8pt, mark=square*, mark options={solid, fill=black, draw=black}, on layer=main, forget plot]
  table[row sep=crcr]{%
0	0\\
1	0\\
2	0\\
3	0\\
4	0\\
5	0\\
6	0\\
7	0\\
8	0\\
9	0\\
10	0\\
11	0\\
12	0\\
13	0\\
14	0\\
15	0\\
16	0\\
17	0\\
18	0\\
19	0\\
20	0\\
21	0\\
22	0\\
23	0\\
24	0\\
25	0\\
26	0\\
27	0\\
28	0\\
29	0\\
30	0\\
31	0\\
32	0\\
33	0\\
34	0\\
35	0\\
36	0\\
0	24\\
1	24\\
2	24\\
3	24\\
4	24\\
5	24\\
6	24\\
7	24\\
8	24\\
9	24\\
10	24\\
11	24\\
12	24\\
13	24\\
14	24\\
15	24\\
16	24\\
17	24\\
18	24\\
19	24\\
20	24\\
21	24\\
22	24\\
23	24\\
24	24\\
25	24\\
26	24\\
27	24\\
28	24\\
29	24\\
30	24\\
31	24\\
32	24\\
33	24\\
34	24\\
35	24\\
36	24\\
0	1\\
0	2\\
0	3\\
0	4\\
0	5\\
0	6\\
0	7\\
0	8\\
0	9\\
0	10\\
0	11\\
0	12\\
0	13\\
0	14\\
0	15\\
0	16\\
0	17\\
0	18\\
0	19\\
0	20\\
0	21\\
0	22\\
0	23\\
36	1\\
36	2\\
36	3\\
36	4\\
36	5\\
36	6\\
36	7\\
36	8\\
36	9\\
36	10\\
36	11\\
36	12\\
36	13\\
36	14\\
36	15\\
36	16\\
36	17\\
36	18\\
36	19\\
36	20\\
36	21\\
36	22\\
36	23\\
};
\addplot [color=mycolor2, only marks, mark size=8.8pt, mark=square*, mark options={solid, fill=black, draw=black}, on layer=main, forget plot]
  table[row sep=crcr]{%
6	1\\
15	1\\
19	1\\
20	1\\
23	1\\
6	2\\
15	2\\
19	2\\
20	2\\
23	2\\
6	3\\
9	3\\
10	3\\
11	3\\
19	3\\
20	3\\
23	3\\
6	4\\
9	4\\
10	4\\
11	4\\
19	4\\
20	4\\
23	4\\
9	5\\
10	5\\
11	5\\
15	5\\
23	5\\
15	6\\
6	7\\
15	7\\
6	8\\
15	8\\
23	8\\
26	8\\
27	8\\
28	8\\
29	8\\
30	8\\
31	8\\
32	8\\
33	8\\
34	8\\
35	8\\
6	9\\
7	9\\
8	9\\
9	9\\
12	9\\
13	9\\
14	9\\
15	9\\
16	9\\
17	9\\
18	9\\
21	9\\
22	9\\
23	9\\
26	9\\
20	12\\
26	12\\
20	13\\
26	13\\
26	14\\
6	16\\
7	16\\
10	16\\
11	16\\
12	16\\
13	16\\
14	16\\
17	16\\
18	16\\
19	16\\
21	16\\
22	16\\
6	17\\
13	17\\
26	17\\
6	18\\
26	18\\
27	18\\
28	18\\
29	18\\
30	18\\
31	18\\
32	18\\
33	18\\
34	18\\
35	18\\
6	19\\
13	20\\
16	20\\
17	20\\
18	20\\
19	20\\
22	20\\
13	21\\
16	21\\
17	21\\
18	21\\
19	21\\
22	21\\
6	22\\
13	22\\
22	22\\
6	23\\
13	23\\
22	23\\
};
\addplot [color=mycolor3, only marks, mark size=8.8pt, mark=square*, mark options={solid, fill=mycolor4, draw=mycolor4}, on layer=axis background, forget plot]
  table[row sep=crcr]{%
26	10\\
26	11\\
26	15\\
26	16\\
};
\addplot [color=white!80!black, only marks, mark size=8.8pt, mark=square*, mark options={solid, fill=white!80!black}, on layer=main, forget plot]
  table[row sep=crcr]{%
9	1\\
10	1\\
11	1\\
9	2\\
10	2\\
11	2\\
15	20\\
15	21\\
15	22\\
20	20\\
20	21\\
16	22\\
17	22\\
18	22\\
19	22\\
20	22\\
17	1\\
17	2\\
17	3\\
17	4\\
17	5\\
17	6\\
18	1\\
18	2\\
18	3\\
18	4\\
18	5\\
18	6\\
19	5\\
19	6\\
20	5\\
20	6\\
21	1\\
21	2\\
21	3\\
21	4\\
21	5\\
21	6\\
22	1\\
22	2\\
22	3\\
22	4\\
22	5\\
22	6\\
15	18\\
16	18\\
17	18\\
18	18\\
19	18\\
20	18\\
15	19\\
16	19\\
17	19\\
18	19\\
19	19\\
20	19\\
19	11\\
19	12\\
19	13\\
19	14\\
20	11\\
20	14\\
21	11\\
21	12\\
21	13\\
21	14\\
};
\addplot [color=mycolor5, only marks, mark size=8pt, mark=*, mark options={solid, fill=red, draw=black}, on layer=axis background, forget plot]
  table[row sep=crcr]{%
30	12\\
};
\addplot [color=mycolor6, line width=1.0pt, on layer=axis foreground, forget plot]
  table[row sep=crcr]{%
10	20\\
11	20\\
11	19\\
11	18\\
12	18\\
13	18\\
14	18\\
15	18\\
15	17\\
16	17\\
17	17\\
18	17\\
19	17\\
20	17\\
20	16\\
20	15\\
21	15\\
22	15\\
23	15\\
24	15\\
25	15\\
26	15\\
27	15\\
27	14\\
27	13\\
28	13\\
29	13\\
30	13\\
30	13\\
30	13\\
30	13\\
30	13\\
30	13\\
30	13\\
30	13\\
};
\addplot [color=mycolor6, only marks, mark size=6.4pt, mark=square*, mark options={solid, fill=mycolor6, draw=black}, forget plot]
  table[row sep=crcr]{%
10	20\\
};
\addplot [color=mycolor6, only marks, mark size=6.4pt, mark=square*, mark options={solid, fill=mycolor6, draw=black}, forget plot]
  table[row sep=crcr]{%
30	13\\
};
\addplot [color=mycolor7, line width=1.0pt, on layer=axis foreground, forget plot]
  table[row sep=crcr]{%
2	16\\
2	15\\
3	15\\
3	14\\
4	14\\
4	13\\
5	13\\
5	12\\
6	12\\
7	12\\
8	12\\
9	12\\
10	12\\
10	11\\
11	11\\
11	10\\
12	10\\
13	10\\
14	10\\
15	10\\
16	10\\
17	10\\
18	10\\
19	10\\
20	10\\
21	10\\
22	10\\
23	10\\
24	10\\
24	11\\
25	11\\
26	11\\
27	11\\
28	11\\
29	11\\
};
\addplot [color=mycolor7, only marks, mark size=6.4pt, mark=square*, mark options={solid, fill=mycolor7, draw=black}, forget plot]
  table[row sep=crcr]{%
2	16\\
};
\addplot [color=mycolor7, only marks, mark size=6.4pt, mark=square*, mark options={solid, fill=mycolor7, draw=black}, forget plot]
  table[row sep=crcr]{%
29	11\\
};
\addplot [color=mycolor9, line width=1.0pt, on layer=axis foreground, forget plot]
  table[row sep=crcr]{%
4	6\\
4	7\\
4	8\\
4	9\\
4	10\\
5	10\\
5	11\\
6	11\\
7	11\\
8	11\\
9	11\\
10	11\\
11	11\\
12	11\\
12	12\\
12	13\\
13	13\\
14	13\\
14	14\\
15	14\\
16	14\\
16	13\\
17	13\\
18	13\\
18	12\\
19	12\\
19	11\\
20	11\\
21	11\\
22	11\\
23	11\\
24	11\\
25	11\\
26	11\\
27	11\\
};
\addplot [color=mycolor9, only marks, mark size=6.4pt, mark=square*, mark options={solid, fill=mycolor9, draw=black}, forget plot]
  table[row sep=crcr]{%
4	6\\
};
\addplot [color=mycolor9, only marks, mark size=6.4pt, mark=square*, mark options={solid, fill=mycolor9, draw=black}, forget plot]
  table[row sep=crcr]{%
27	11\\
};
\addplot [color=mycolor11, line width=1.0pt, on layer=axis foreground, forget plot]
  table[row sep=crcr]{%
12	2\\
12	3\\
13	3\\
14	3\\
14	4\\
15	4\\
16	4\\
16	5\\
17	5\\
18	5\\
19	5\\
19	6\\
20	6\\
20	7\\
20	8\\
20	9\\
20	10\\
21	10\\
21	11\\
22	11\\
22	12\\
23	12\\
23	13\\
24	13\\
24	14\\
25	14\\
25	15\\
26	15\\
27	15\\
28	15\\
29	15\\
29	14\\
29	13\\
29	13\\
29	13\\
};
\addplot [color=mycolor2, only marks, mark size=6.4pt, mark=square*, mark options={solid, fill=mycolor11, draw=black}, forget plot]
  table[row sep=crcr]{%
12	2\\
};
\addplot [color=mycolor3, only marks, mark size=6.4pt, mark=square*, mark options={solid, fill=mycolor11, draw=black}, forget plot]
  table[row sep=crcr]{%
29	13\\
};
\end{axis}

\end{tikzpicture}%
\end{minipage}
\hfill
\begin{minipage}[c]{0.48\textwidth}
\centering
%
%
\definecolor{mycolor1}{rgb}{0.00000,0.44700,0.74100}%
\definecolor{mycolor2}{rgb}{0.85000,0.32500,0.09800}%
\definecolor{mycolor3}{rgb}{0.92900,0.69400,0.12500}%
\definecolor{mycolor4}{rgb}{1.00000,1.00000,0.00000}%
\definecolor{mycolor5}{rgb}{0,1,0}%
\definecolor{mycolor6}{rgb}{0.180, 0.545, 0.341}%
\definecolor{mycolor7}{rgb}{0.30196,0.74510,0.93333}%
\definecolor{mycolor8}{rgb}{0.49400,0.18400,0.55600}%
\definecolor{mycolor9}{rgb}{0.855, 0.439, 0.839}%
\definecolor{mycolor10}{rgb}{0.992, 0.737, 0.706}%
\definecolor{mycolor11}{rgb}{0, 0, 1}%
%
\begin{tikzpicture}[scale=0.32]
\tikzset{every node/.style={font=\small}}
\pgfplotsset{set layers}

\begin{axis}[%
width=8.427in,
height=5.694in,
at={(0.612in,0.864in)},
scale only axis,
xmin=-0.539055724707299,
xmax=36.4609442752927,
xtick={0,4,8,12,16,20,24,28,32,36},
tick label style={font=\small},
xlabel style={font=\color{white!15!black}, yshift=-10pt},
xlabel={x [m]}, 
xlabel style={font=\small},
ymin=-0.563745259817576,
ymax=24.4362547401824,
ytick={ 0,  4,  8, 12, 16, 20, 24, 28},
ylabel style={font=\color{white!15!black}},
ylabel={y [m]}, 
ylabel style={font=\small, yshift=18pt},
axis background/.style={fill=white}
]
\addplot [color=mycolor1, only marks, mark size=8.8pt, mark=square*, mark options={solid, fill=black, draw=black}, forget plot]
  table[row sep=crcr]{%
0	0\\
1	0\\
2	0\\
3	0\\
4	0\\
5	0\\
6	0\\
7	0\\
8	0\\
9	0\\
10	0\\
11	0\\
12	0\\
13	0\\
14	0\\
15	0\\
16	0\\
17	0\\
18	0\\
19	0\\
20	0\\
21	0\\
22	0\\
23	0\\
24	0\\
25	0\\
26	0\\
27	0\\
28	0\\
29	0\\
30	0\\
31	0\\
32	0\\
33	0\\
34	0\\
35	0\\
36	0\\
0	24\\
1	24\\
2	24\\
3	24\\
4	24\\
5	24\\
6	24\\
7	24\\
8	24\\
9	24\\
10	24\\
11	24\\
12	24\\
13	24\\
14	24\\
15	24\\
16	24\\
17	24\\
18	24\\
19	24\\
20	24\\
21	24\\
22	24\\
23	24\\
24	24\\
25	24\\
26	24\\
27	24\\
28	24\\
29	24\\
30	24\\
31	24\\
32	24\\
33	24\\
34	24\\
35	24\\
36	24\\
0	1\\
0	2\\
0	3\\
0	4\\
0	5\\
0	6\\
0	7\\
0	8\\
0	9\\
0	10\\
0	11\\
0	12\\
0	13\\
0	14\\
0	15\\
0	16\\
0	17\\
0	18\\
0	19\\
0	20\\
0	21\\
0	22\\
0	23\\
36	1\\
36	2\\
36	3\\
36	4\\
36	5\\
36	6\\
36	7\\
36	8\\
36	9\\
36	10\\
36	11\\
36	12\\
36	13\\
36	14\\
36	15\\
36	16\\
36	17\\
36	18\\
36	19\\
36	20\\
36	21\\
36	22\\
36	23\\
};
\addplot [color=mycolor2, only marks, mark size=8.8pt, mark=square*, mark options={solid, fill=black, draw=black}, forget plot]
  table[row sep=crcr]{%
6	1\\
15	1\\
19	1\\
20	1\\
23	1\\
6	2\\
15	2\\
19	2\\
20	2\\
23	2\\
6	3\\
9	3\\
10	3\\
11	3\\
19	3\\
20	3\\
23	3\\
6	4\\
9	4\\
10	4\\
11	4\\
19	4\\
20	4\\
23	4\\
9	5\\
10	5\\
11	5\\
15	5\\
23	5\\
15	6\\
6	7\\
15	7\\
6	8\\
15	8\\
23	8\\
26	8\\
27	8\\
28	8\\
29	8\\
30	8\\
31	8\\
32	8\\
33	8\\
34	8\\
35	8\\
6	9\\
7	9\\
8	9\\
9	9\\
12	9\\
13	9\\
14	9\\
15	9\\
16	9\\
17	9\\
18	9\\
21	9\\
22	9\\
23	9\\
26	9\\
20	12\\
26	12\\
20	13\\
26	13\\
26	14\\
6	16\\
7	16\\
10	16\\
11	16\\
12	16\\
13	16\\
14	16\\
17	16\\
18	16\\
19	16\\
21	16\\
22	16\\
6	17\\
13	17\\
26	17\\
6	18\\
26	18\\
27	18\\
28	18\\
29	18\\
30	18\\
31	18\\
32	18\\
33	18\\
34	18\\
35	18\\
6	19\\
13	20\\
16	20\\
17	20\\
18	20\\
19	20\\
22	20\\
13	21\\
16	21\\
17	21\\
18	21\\
19	21\\
22	21\\
6	22\\
13	22\\
22	22\\
6	23\\
13	23\\
22	23\\
};
\addplot [color=mycolor3, only marks, mark size=8.8pt, mark=square*, mark options={solid, fill=mycolor4, draw=mycolor4}, forget plot]
  table[row sep=crcr]{%
26	10\\
26	11\\
26	15\\
26	16\\
};
\addplot [color=white!80!black, only marks, mark size=8.8pt, mark=square*, mark options={solid, fill=white!80!black}, forget plot]
  table[row sep=crcr]{%
9	1\\
10	1\\
11	1\\
9	2\\
10	2\\
11	2\\
15	20\\
15	21\\
15	22\\
20	20\\
20	21\\
16	22\\
17	22\\
18	22\\
19	22\\
20	22\\
17	1\\
17	2\\
17	3\\
17	4\\
17	5\\
17	6\\
18	1\\
18	2\\
18	3\\
18	4\\
18	5\\
18	6\\
19	5\\
19	6\\
20	5\\
20	6\\
21	1\\
21	2\\
21	3\\
21	4\\
21	5\\
21	6\\
22	1\\
22	2\\
22	3\\
22	4\\
22	5\\
22	6\\
15	18\\
16	18\\
17	18\\
18	18\\
19	18\\
20	18\\
15	19\\
16	19\\
17	19\\
18	19\\
19	19\\
20	19\\
19	11\\
19	12\\
19	13\\
19	14\\
20	11\\
20	14\\
21	11\\
21	12\\
21	13\\
21	14\\
};
\addplot [color=mycolor5, only marks, mark size=8pt, mark=*, mark options={solid, fill=red, draw=black}, on layer=axis background, forget plot]
  table[row sep=crcr]{%
30	12\\
};
\addplot [color=mycolor6, line width=1.0pt, on layer=axis foreground, forget plot]
  table[row sep=crcr]{%
10	20\\
11	20\\
11	19\\
12	19\\
13	19\\
14	19\\
14	18\\
14	17\\
15	17\\
15	16\\
16	16\\
16	15\\
16	14\\
17	14\\
17	15\\
18	15\\
19	15\\
20	15\\
21	15\\
22	15\\
23	15\\
24	15\\
25	15\\
26	15\\
26	16\\
27	16\\
27	15\\
27	14\\
27	15\\
27	14\\
27	15\\
27	14\\
27	15\\
27	14\\
27	15\\
};
\addplot [color=mycolor6, only marks, mark size=6.4pt, mark=square*, mark options={solid, fill=mycolor6, draw=black}, forget plot]
  table[row sep=crcr]{%
10	20\\
};
\addplot [color=mycolor1, only marks, mark size=6.4pt, mark=square*, mark options={solid, fill=mycolor6, draw=black}, forget plot]
  table[row sep=crcr]{%
27	15\\
};
\addplot [color=mycolor7, line width=1.0pt, on layer=axis foreground, forget plot]
  table[row sep=crcr]{%
2	16\\
2	15\\
3	15\\
4	15\\
5	15\\
6	15\\
7	15\\
8	15\\
8	16\\
9	16\\
9	17\\
9	18\\
10	18\\
11	18\\
12	18\\
13	18\\
14	18\\
14	17\\
15	17\\
16	17\\
17	17\\
18	17\\
19	17\\
20	17\\
21	17\\
22	17\\
23	17\\
24	17\\
25	17\\
25	16\\
26	16\\
27	16\\
27	16\\
27	15\\
28	15\\
};
\addplot [color=mycolor7, only marks, mark size=6.4pt, mark=square*, mark options={solid, fill=mycolor7, draw=black}, forget plot]
  table[row sep=crcr]{%
2	16\\
};
\addplot [color=mycolor7, only marks, mark size=6.4pt, mark=square*, mark options={solid, fill=mycolor7, draw=black}, forget plot]
  table[row sep=crcr]{%
28	15\\
};
\addplot [color=mycolor9, line width=1.0pt, on layer=axis foreground, forget plot]
  table[row sep=crcr]{%
4	6\\
4	7\\
5	7\\
5	8\\
5	9\\
5	10\\
6	10\\
7	10\\
8	10\\
8	11\\
9	11\\
10	11\\
11	11\\
11	10\\
12	10\\
13	10\\
14	10\\
15	10\\
16	10\\
17	10\\
18	10\\
19	10\\
20	10\\
21	10\\
22	10\\
23	10\\
24	10\\
25	10\\
25	11\\
26	11\\
27	11\\
27	12\\
28	12\\
29	12\\
29	12\\
};
\addplot [color=mycolor9, only marks, mark size=6.4pt, mark=square*, mark options={solid, fill=mycolor9, draw=black}, forget plot]
  table[row sep=crcr]{%
4	6\\
};
\addplot [color=mycolor9, only marks, mark size=6.4pt, mark=square*, mark options={solid, fill=mycolor9, draw=black}, forget plot]
  table[row sep=crcr]{%
29	12\\
};
\addplot [color=mycolor11, line width=1.0pt, on layer=axis foreground, forget plot]
  table[row sep=crcr]{%
12	2\\
13	2\\
13	3\\
14	3\\
15	3\\
16	3\\
16	4\\
16	5\\
16	6\\
16	7\\
16	8\\
17	8\\
18	8\\
19	8\\
19	9\\
20	9\\
20	10\\
21	10\\
22	10\\
23	10\\
24	10\\
25	10\\
26	10\\
26	11\\
27	11\\
28	11\\
29	11\\
29	10\\
29	11\\
29	10\\
29	11\\
29	10\\
29	11\\
29	10\\
29	11\\
};
\addplot [color=mycolor11, only marks, mark size=6.4pt, mark=square*, mark options={solid, fill=mycolor11, draw=black}, forget plot]
  table[row sep=crcr]{%
12	2\\
};
\addplot [color=mycolor11, fill opacity=1, only marks, mark size=6.4pt, mark=square*, mark options={solid, fill=mycolor11, draw=black}, forget plot]
  table[row sep=crcr]{%
29	11\\
};
\end{axis}

\end{tikzpicture}%
\end{minipage}
\caption{Final trajectories for the instrumental-only (left) and Pavlovian (right) agents.}
\label{fig:finaltraj}
\end{figure*}

Then, Figure~\ref{fig:pavlovianep1} illustrates the evolution of the Pavlovian state value for the four agents across training, more specifically for episode $1$ (top row) and $400$ (bottom row). Unlike standard \ac{RL} value functions that typically reflect distance or progress toward a goal, the Pavlovian value forms a “risk–reward potential field’’ shaped by environmental radio cues. Distinct local maxima emerge around gate locations, while strong minima appear in \ac{GPS}-denied regions. These potentials act as intrinsic shaping rewards: for example, the negative values associated with \ac{GPS}-denied zones suppress actions that would lead toward them, effectively pruning the instrumental learner’s search space.


Figure \ref{fig:finaltraj} further highlights the effect of Pavlovian conditioning on the final trajectories. Under instrumental-only learning, agents find viable routes but follow longer, loosely structured paths that sometimes approach \ac{GPS}-denied areas. This reflects their reliance on uninformed trial-and-error exploration, which requires extensive sampling to separate useful corridors from deceptive ones, leading to persistent inefficiencies. In contrast, Pavlovian learners trajectories are more direct, maintain safe margins around GPS-denied regions, and exploit LOS corridors more effectively. This reflects that Pavlovian conditioning not only accelerates learning but also helps to achieve final policies that are more responsive to the surrounding environment.


\section{Conclusions \& Future Perspectives}
\label{sec:conclusions}
This work introduced a human-inspired \ac{RL} framework that integrates Pavlovian and instrumental learning processes into autonomous digital agents. By translating fundamental mechanisms of human cognition, such as cue-driven Pavlovian learning, into the design of digital agents, we demonstrated through a simple case study that radio cues can serve as predictive signals that accelerate learning in complex environments, outperforming performance of instrumental-solo learners, and highlighting the potentialities offered by human cognitive principles. 

Looking ahead, several research directions emerge naturally from this work. First, collective learning mechanisms represent a compelling extension: just as humans share knowledge through communication, autonomous agents could exchange cue–outcome associations or value structures, enabling efficient transfer learning among agents. This may pave the way for emergent collective behaviors, in which agents adopting \ac{ST} or \ac{GT} attitudes, self-organize into complementary functional roles.
Indeed, \acp{ST} are agents who tend to respond more strongly to Pavlovian cues and rely on model-free (trial-and-error) strategies, while \acp{GT} emphasize outcome prediction and model-based strategies \cite{dayan2002reward}. 
Second, the implementation of more sophisticated arbitration strategies to balance model-based and -free systems could further enhance flexibility, especially in rapidly changing environments. Third, the extraction of radio cues from semantic radio maps remains an open research problem and therefore requires careful investigation. Finally, integrating higher-level cognitive skills, such as social inference, trust evaluation, and cooperative reasoning, could bring autonomous systems even closer to human intelligence.

\bibliographystyle{IEEEtran}
\bibliography{Biblio}

@IEEEtranBSTCTL{IEEEexample:BSTcontrol,
CTLuse_forced_etal = "yes",
CTLmax_names_forced_etal = "3",
CTLnames_show_etal = "1",
}

@article{SaeEtAl:C26,
  author  = { Saeidi, Mehrdad and Shan, Jingfeng and  Li Pira, Giorgio and Starita, Francesca and Guidi, Francesco and Guerra, Anna},
  title   = {PAVLOVIAN-INSPIRED CUE-OUTCOME ASSOCIATION FOR AUTONOMOUS
AGENT NAVIGATION},
  journal = {to Appear in ICASSP 2026},
}

@article{dalbagno2025learning,
  title={Learning the time of pain in the human motor system},
  author={Dalbagno, Daniela and Betti, Sonia and Garofalo, Sara and Mannari, Vanessa and Di Pellegrino, Giuseppe and Starita, Francesca},
  journal={Pain},
  volume={166},
  number={12},
  pages={e715--e731},
  year={2025},
  publisher={LWW}
}

@article{Dorigo2021,
  author  = {Dorigo, Marco and Theraulaz, Guy and Trianni, Vito},
  title   = {Swarm Robotics: Past, Present, and Future [Point of View]},
  journal = {Proc. IEEE},
  volume  = {109},
  number  = {7},
  pages   = {1152--1165},
  year    = {2021},
  doi     = {10.1109/JPROC.2021.3080080}
}

@article{HayEtAl:J2,
  title={Cognitive control},
  author={Haykin, Simon and Fatemi, Mehdi and Setoodeh, Peyman and Xue, Yanbo},
  journal={Proc. IEEE},
  volume={100},
  number={12},
  pages={3156--3169},
  year={2012},
}

@article{Guerra2020a,
  author  = {Guerra, Anna and Dardari, Davide and Djuri{\'c}, Petar M.},
  title   = {Dynamic radar network of {UAV}s: A joint navigation and tracking approach},
  journal = {IEEE Access},
  volume  = {8},
  pages   = {116454--116469},
  year    = {2020},
  doi     = {10.1109/ACCESS.2020.3004533}
}

@article{Guerra2020b,
  author  = {Guerra, Antonio and Dardari, Davide and Muntean, Gabriel-Mugurel and Djuri{\'c}, Petar M.},
  title   = {Dynamic radar networks of {UAV}s: A tutorial overview and tracking performance comparison with terrestrial radar networks},
  journal = {IEEE Veh. Technol. Mag.},
  volume  = {15},
  number  = {2},
  pages   = {113--120},
  year    = {2020},
  doi     = {10.1109/MVT.2020.2988450}
}

@article{dabney2020distributional,
  title={A distributional code for value in dopamine-based reinforcement learning},
  author={Dabney, Will and Kurth-Nelson, Zeb and Uchida, Naoshige and Starkweather, Clara Kwon and Hassabis, Demis and Munos, R{\'e}mi and Botvinick, Matthew},
  journal={Nature},
  volume={577},
  number={7792},
  pages={671--675},
  year={2020},
  publisher={Nature Publishing Group UK London}
}

@inproceedings{Guerra2021,
  author    = {Guerra, Anna and Guidi, Francesco and Dardari, Davide and Djuri{\'c}, Petar M.},
  title     = {Real-time learning for {TH}z radar mapping and {UAV} control},
  booktitle = {Proc. IEEE Int. Conf. on Autonomous Syst. (ICAS)},
  year      = {2021},
  pages     = {1--6},
}

@article{Hassabis2017,
  author  = {Hassabis, Demis and Kumaran, Dharshan and Summerfield, Christopher and Botvinick, Matthew},
  title   = {Neuroscience-inspired artificial intelligence},
  journal = {Neuron},
  volume  = {95},
  number  = {2},
  pages   = {245--258},
  year    = {2017},
  doi     = {10.1016/j.neuron.2017.06.011}
}

@article{Hodge2021,
  author  = {Hodge, Victoria J. and Hawkins, Robert and Alexander, Rob},
  title   = {Deep reinforcement learning for drone navigation using sensor data},
  journal = {Neural Computing and Applications},
  volume  = {33},
  number  = {6},
  pages   = {2015--2033},
  year    = {2021},
  doi     = {10.1007/s00521-020-05017-2}
}

@article{Kalia2008,
  author  = {Kalia, Amy A. and Legge, Gordon E. and Giudice, Nicholas A.},
  title   = {Learning building layouts with non-geometric visual information: The effects of visual impairment and age},
  journal = {Perception},
  volume  = {37},
  number  = {11},
  pages   = {1677--1699},
  year    = {2008},
  doi     = {10.1068/p5972}
}

@article{Meyer2015,
  author  = {Meyer, Florian and Hlinka, Ondrej and Wymeersch, Henk and Hlawatsch, Franz and Djuric, Petar M.},
  title   = {Distributed estimation with information-seeking control in agent networks},
  journal = {IEEE J. Sel. Areas Commun.},
  volume  = {33},
  number  = {11},
  pages   = {2439--2456},
  year    = {2015},
  doi     = {10.1109/JSAC.2015.2430293}
}

@article{Na2022,
  author  = {Na, Seungmin and Cho, Jaeho and Kim, Hyunsoo and Lee, Sangwoo},
  title   = {Bio-inspired Collision Avoidance in Swarm Systems via Deep Reinforcement Learning},
  journal = {IEEE Trans. Veh. Technol.},
  year    = {2022},
  doi     = {10.1109/TVT.2022.3174593}
}

@article{fontanesi2025deep,
  title={A deep-{NN} beamforming approach for dual function radar-communication {TH}z {UAV}},
  author={Fontanesi, Gianluca and Guerra, Anna and Guidi, Francesco and V{\'a}squez-Peralvo, Juan A and Shlezinger, Nir and Zanella, Alberto and Lagunas, Eva and Chatzinotas, Symeon and Dardari, Davide and Djuri{\'c}, Petar M},
  journal={IEEE Trans. Veh. Technol.},
year={2025},
  volume={74},
  number={1},
  pages={746-760},
}

@article{Jaiswal2022,
  author  = {Jaiswal, Saurabh and Sidhanta, Shibashis},
  title   = {Toward a smart multi-unmanned aerial vehicle system},
  journal = {IEEE Potentials},
  volume  = {41},
  number  = {1},
  pages   = {22--25},
  year    = {2022},
  doi     = {10.1109/MPOT.2021.3079944}
}

@inproceedings{Testi2020,
  author    = {Testi, Enrico and Favarelli, Elisa and Giorgetti, Andrea},
  title     = {Reinforcement learning for connected autonomous vehicle localization via {UAV}s},
  booktitle = {Proc. IEEE Int. Workshop on Metrology for Agriculture and Forestry (MetroAgriFor)},
  year      = {2020},
  pages     = {101--106},
  doi       = {10.1109/MetroAgriFor49836.2020.9318860}
}

@article{Wu2021,
  author  = {Wu, Jiehong and Song, Chengxin and Ma, Jian and Wu, Jinsong and Han, Guangjie},
  title   = {Reinforcement Learning and Particle Swarm Optimization Supporting Real‑Time Rescue Assignments for Multiple Autonomous Underwater Vehicles},
  journal = {IEEE Trans. Intell. Transp. Syst.},
  volume  = {23},
  number  = {8},
  pages   = {6807--6820},
  year    = {2021},
  doi     = {10.1109/TITS.2021.3062500}
}

@article{Wen2020,
  author  = {Wen, Shuhuan and Zhao, Yanfang and Yuan, Xiao and Wang, Zongtao and Zhang, Dan and Manfredi, Luigi},
  title   = {Path planning for active {SLAM} based on deep reinforcement learning under unknown environments},
  journal = {Intelligent Service Robotics},
  volume  = {13},
  number  = {2},
  pages   = {263--272},
  year    = {2020},
  doi     = {10.1007/s11370-019-00310-w}
}

@article{EuropeanCommissionJRC2025,
  author       = {{European Commission, Joint Research Centre}},
  title        = {Eyes on the Future – Signals from Recent Reports on Emerging Technologies and Breakthrough Innovations to Support European Innovation Council Strategic Intelligence},
  institution  = {Publications Office of the European Union},
  year={2025},
   volume={3},
  url={https://data.europa.eu/doi/10.2760/1104283},
}

@article{chen2024double,
  title={Double Actor-Critic with {TD} Error-Driven Regularization in Reinforcement Learning},
  author={Chen, Haohui and Chen, Zhiyong and Liu, Aoxiang and Fang, Wentuo},
  journal={arXiv preprint arXiv:2409.19231},
  year={2024}
}

@article{starita2023threat,
  title={Threat learning in space: how stimulus-outcome spatial compatibility modulates conditioned skin conductance response},
  author={Starita, F and Stussi, Y and Garofalo, S and di Pellegrino, G},
  journal={Int. J. Psychophysiology},
  volume={190},
  pages={30--41},
  year={2023},
  publisher={Elsevier}
}

@book{sutton1998reinforcement,
  title={Reinforcement learning: An introduction},
  author={Sutton, Richard S and Barto, Andrew G and others},
  volume={1},
  number={1},
  year={1998},
  publisher={MIT press Cambridge}
}

@article{campo2021collective,
  title={Collective Memory: {T}ransposing Pavlov’s Experiment to Robot Swarms},
  author={Campo, Alexandre and Nicolis, Stamatios C and Deneubourg, Jean-Louis},
  journal={Applied Sciences},
  volume={11},
  number={6},
  pages={2632},
  year={2021},
  publisher={MDPI}
}

@article{zhang2024cognitive,
  title={Cognitive control architecture for the practical realization of {UAV} collision avoidance},
  author={Zhang, Qirui and Wei, Ruixuan and Huang, Songlin},
  journal={Sensors},
  volume={24},
  number={9},
  pages={2790},
  year={2024},
  publisher={MDPI}
}

@article{haykin2006cognitive,
  title={Cognitive radar: a way of the future},
  author={Haykin, Simon},
  journal={IEEE Signal Process. Mag.},
  volume={23},
  number={1},
  pages={30--40},
  year={2006},
}

@article{schultz1997neural,
  title={A neural substrate of prediction and reward},
  author={Schultz, Wolfram and Dayan, Peter and Montague, P. Read},
  journal={Science},
  volume={275},
  number={5306},
  pages={1593--1599},
  year={1997},
  publisher={American Association for the Advancement of Science},
  doi={10.1126/science.275.5306.1593}
}

@article{dayan2002reward,
  title={Reward, motivation, and reinforcement learning},
  author={Dayan, Peter and Balleine, Bernard W.},
  journal={Neuron},
  volume={36},
  number={2},
  pages={285--298},
  year={2002},
  publisher={Elsevier},
  doi={10.1016/S0896-6273(02)00963-7}
}

@article{pool2019behavioural,
  title={Behavioural evidence for parallel outcome-sensitive and outcome-insensitive {P}avlovian learning systems in humans},
  author={Pool, Eva R and Pauli, Wolfgang M and Kress, Carolina S and O’Doherty, John P},
  journal={Nature human behaviour},
  volume={3},
  number={3},
  pages={284--296},
  year={2019},
  publisher={Nature Publishing Group UK London}
}

@article{dayan2014model,
  title={Model-based and model-free {P}avlovian reward learning: {R}evaluation, revision, and revelation},
  author={Dayan, Peter and Berridge, Kent C},
  journal={Cognitive, Affective, \& Behavioral Neuroscience},
  volume={14},
  pages={473--492},
  year={2014},
  publisher={Springer}
}

@article{geerts2020general,
  title={A general model of hippocampal and dorsal striatal learning and decision making},
  author={Geerts, Jesse P and Chersi, Fabian and Stachenfeld, Kimberly L and Burgess, Neil},
  journal={Proc. of the National Academy of Sciences},
  volume={117},
  number={49},
  pages={31427--31437},
  year={2020},
  publisher={National Academy of Sciences}
}

@article{mahajan2024balancing,
  title={Balancing safety and efficiency in human decision making},
  author={Mahajan, Pranav and Tong, Shuangyi and Lee, Sang Wan and Seymour, Ben},
  journal={bioRxiv},
  pages={2024--01},
  year={2024},
  publisher={Cold Spring Harbor Laboratory}
}

@inproceedings{elfwing2017parallel,
  title={Parallel reward and punishment control in humans and robots: Safe reinforcement learning using the MaxPain algorithm},
  author={Elfwing, Stefan and Seymour, Ben},
  booktitle={Proc. Joint IEEE Int. Conf. on Development and Learning and Epigenetic Robot. (ICDL-EpiRob)},
  pages={140--147},
  year={2017},
}

@article{wang2021modular,
  title={Modular deep reinforcement learning from reward and punishment for robot navigation},
  author={Wang, Jiexin and Elfwing, Stefan and Uchibe, Eiji},
  journal={Neural Networks},
  volume={135},
  pages={115--126},
  year={2021},
  publisher={Elsevier}
}

@article{lee2019decision,
  title={Decision-making in brains and robots—The case for an interdisciplinary approach},
  author={Lee, Sang Wan and Seymour, Ben},
  journal={Current Opinion in Behavioral Sciences},
  volume={26},
  pages={137--145},
  year={2019},
  publisher={Elsevier}
}

@article{haykin2012cognitive,
  title={Cognitive radar: {S}tep toward bridging the gap between neuroscience and engineering},
  author={Haykin, Simon and Xue, Yanbo and Setoodeh, Peyman},
  journal={Proc.  {IEEE}},
  volume={100},
  number={11},
  pages={3102--3130},
  year={2012},
  publisher={IEEE}
}

@article{starita2022pavlovian,
  title={Pavlovian threat learning shapes the kinematics of action},
  author={Starita, Francesca and Garofalo, Sara and Dalbagno, Daniela and Degni, Luigi AE and Di Pellegrino, Giuseppe},
  journal={Frontiers in Psychology},
  volume={13},
  pages={1005656},
  year={2022},
  publisher={Frontiers Media SA}
}

@article{glascher2010states,
  title={States versus rewards: dissociable neural prediction error signals underlying model-based and model-free reinforcement learning},
  author={Gl{\"a}scher, Jan and Daw, Nathaniel and Dayan, Peter and O'Doherty, John P},
  journal={Neuron},
  volume={66},
  number={4},
  pages={585--595},
  year={2010},
  publisher={Elsevier}
}

@article{delamater2007learning,
  title={Learning about multiple attributes of reward in Pavlovian conditioning},
  author={Delamater, Andrew R and Oakeshott, Stephen},
  journal={Annals of the New York Academy of Sciences},
  volume={1104},
  number={1},
  pages={1--20},
  year={2007},
  publisher={Wiley Online Library}
}

@article{guerra2022networks,
  title={Networks of {UAV}s of low complexity for time-critical localization},
  author={Guerra, Anna and Guidi, Francesco and Dardari, Davide and Djuri{\'c}, Petar M},
  journal={IEEE Aerosp. Electron. Syst. Mag.},
  volume={37},
  number={10},
  pages={22--38},
  year={2022},
  publisher={IEEE}
}

@article{moore1993prioritized,
  title        = {Prioritized Sweeping: Reinforcement Learning with Less Data and Less Real Time},
  author       = {Moore, Andrew W. and Atkeson, Christopher G.},
  journal      = {Machine Learning},
  volume       = {13},
  number       = {1},
  pages        = {103--130},
  year         = {1993}
}

@book{anderson2007optimal,
  title={Optimal Control: Linear Quadratic Methods},
  author={Anderson, Brian D.O. and Moore, John B.},
  year={2007},
  publisher={Dover Publications}
}

@book{rawlings2009model,
  title={Model Predictive Control: Theory and Design},
  author={Rawlings, James B. and Mayne, David Q.},
  year={2009},
  publisher={Nob Hill Publishing}
}

@article{schrittwieser2020mastering,
  title   = {Mastering Atari, Go, Chess and Shogi by Planning with a Learned Model},
  author  = {Schrittwieser, Julian and Antonoglou, Ioannis and Hubert, Thomas and Simonyan, Karen and Sifre, Laurent and Schmitt, Simon and Guez, Arthur and Lockhart, Edward and Hassabis, Demis and Graepel, Thore and Lillicrap, Timothy and Silver, David},
  journal = {Nature},
  volume  = {588},
  number  = {7839},
  pages   = {604--609},
  year    = {2020},
  doi     = {10.1038/s41586-020-03051-4}
}

@inproceedings{hafner2019learning,
  title     = {Learning Latent Dynamics for Planning from Pixels},
  author    = {Hafner, Danijar and Lillicrap, Timothy and Fischer, Ian and Villegas, Ruben and Ha, David and Lee, Honglak and Davidson, James},
  booktitle = {Proc. 36th Int. Conf. on Machine Learning (ICML)},
  pages     = {2555--2565},
  year      = {2019}
}

@inproceedings{janner2019mbpo,
  title        = {When to Trust Your Model: Model-Based Policy Optimization},
  author       = {Janner, Michael and Fu, Justin and Zhang, Marvin and Levine, Sergey},
  booktitle    = {Advances in Neural Information Processing Systems 32 (NeurIPS)},
  year         = {2019}
}

@inproceedings{sutton1990dyna,
  author       = {Sutton, Richard S.},
  title        = {Integrated Architectures for Learning, Planning, and Reacting Based on Approximating Dynamic Programming},
  booktitle    = {Proc.\ 7th Int.\ Conf.\ on Machine Learning (ICML)},
  pages        = {216--224},
  year         = {1990},
  publisher    = {Morgan Kaufmann},
  address      = {San Mateo, CA}
}

@article{lee2014neural,
  title={Neural computations underlying arbitration between model-based and model-free learning},
  author={Lee, Sang Wan and Shimojo, Shinsuke and O’doherty, John P},
  journal={Neuron},
  volume={81},
  number={3},
  pages={687--699},
  year={2014},
  publisher={Elsevier}
}

@article{o2015structure,
  title={The structure of reinforcement-learning mechanisms in the human brain},
  author={O’Doherty, John P and Lee, Sang Wan and McNamee, Daniel},
  journal={Current Opinion in Behavioral Sciences},
  volume={1},
  pages={94--100},
  year={2015},
  publisher={Elsevier}
}

@article{mnih2015human,
  title={Human-level control through deep reinforcement learning},
  author={Mnih, Volodymyr and Kavukcuoglu, Koray and Silver, David and Rusu, Andrei A and Veness, Joel and Bellemare, Marc G and Graves, Alex and Riedmiller, Martin and Fidjeland, Andreas K and Ostrovski, Georg and others},
  journal={nature},
  volume={518},
  number={7540},
  pages={529--533},
  year={2015},
  publisher={Nature Publishing Group}
}

\end{document}